\documentstyle[prl,aps]{revtex}
\begin{document}
\draft
\tighten

\title{
Delocalization of Flux Lines from Extended Defects by Bulk Randomness}
\author{Leon Balents}
\address{Department of Physics, Harvard University, Cambridge, MA 02138}

\author{Mehran Kardar}
\address{Physics Department, Massachusetts Institute of Technology,
Cambridge, MA 02139}
\date{\today}
\maketitle
\begin{abstract}
We study the delocalization by bulk randomness of a single flux line
(FL) from an extended defect, such as a columnar pin or twin plane.
In three dimensions, the FL is always bound to a planar defect, while
there is an unpinning transition from a columnar pin.  Transfer matrix
simulations confirm this picture, and indicate that the divergence of
the localization length from the columnar defect is governed by a
liberation exponent $\nu_\perp =1.3 \pm 0.6$, for which a ``mean-field''
estimate gives $\nu_\perp \approx 0.78$. The results, and their
extensions, are compared to other theories.  The effects may be
observable in thin samples close to $H_{c1}$.
\end{abstract}
\narrowtext
\twocolumn

It is now well known that fluctuations can drastically change the
nature of the Abrikosov phase of type II superconductors.  Thermal
disorder can melt the vortex lattice, forming a flux liquid at both
low and high vortex densities\cite{DRN}. Quenched randomness
also leads to new behavior\cite{LO}, which depends upon its degree of
correlation.  Point defects, such as vacancies or interstitials,
encourage line wandering, and may lead to the formation of a glassy
phase with a non-zero critical current, called a vortex
glass\cite{VG}.  In the extreme low density limit of a single flux line
(FL), analytical \cite{KZ} and numerical
\cite{Numerics} work has shown that this  is indeed the case, though the
existence of the vortex glass away from $H_{c1}$ is still unclear.
More recently, the experimental creation of ``columnar,'' or linear,
pinning sites \cite{Civale} has inspired theoretical treatment of the
resulting ``Bose glass'' phase\cite{NV}. Twin planes, planar defects
which occur naturally as a type of grain boundary, may lead to a
different phase\cite{NV}.

In this letter, we study the
behavior of an individual flux line in the presence of
{\sl both} point disorder and a single extended defect
(e.g.\ a columnar pin; see Fig.\ \ref{fig1}).
 The
free energy of a $d$-dimensional FL of length $L$
interacting with an $n$ dimensional extended defect is
\begin{equation}
F = \int_0^L \!\! dz \!\left\{ {\tilde{\epsilon} \over 2}\left| {{d
{\bf x}(z)} \over {dz}}\right|^2 \!\!+ \! V_{\rm R}({\bf x}(z),z)\!
-\! \Delta \delta({\bf x}_\perp(z))
\right\}\! ,
\label{eFreeenergy}
\end{equation}
where ${\bf x}(z) \in \Re^{d-1}$ is the displacement of the flux line,
${\bf x}_\perp(z) \in \Re^{d-n}$ is the component perpendicular to the
defect, $\tilde{\epsilon}$ is the stiffness, $\delta({\bf
x}_\perp)$ is a short-range potential concentrated near ${\bf x}_\perp
= 0$, and $V_{\rm R}({\bf x},z)$ is a random potential representing
the point disorder, with
\begin{equation}
\langle V_{\rm R}({\bf x},z)V_{\rm R}({\bf x}',z') \rangle = \sigma^2
\delta({\bf x-
x'})\delta(z-z').
\end{equation}
For $n=1$ or $n=2$, this model describes pinning by a columnar or a
planar defect, respectively.

In $d\leq3$, the wanderings of a single flux line
(FL) are controlled by the random potential due to impurities.
While in higher dimensions, a line freely fluctuating due to thermal
effects is stable to weak randomness\cite{KZ}, in this paper we
consider mainly situations where the randomness is strong enough
to be relevant. In such cases, the behavior at long length scales
is dominated by a zero temperature fixed point\cite{VG}. However,
even at $T=0$, there can be a non-trivial transition between flux
lines localized or free from the extended defect\cite{Kreplica}.
The zero temperature phase transition also governs the
singular behavior at finite
temperatures. The relevance of disorder, and the dominance of zero
temperature fixed points simplifies certain aspects of the calculation.
The partition sum over all thermally fluctuating configurations is
dominated by the optimal path which minimizes Eq.\ref{eFreeenergy},
now regarded as an {\sl energy}.

The behavior of the FL in the absence of the extended defect,
$\Delta = 0$, is well-understood from extensive theoretical and
numerical work\cite{KZ,FH,Numerics}. The energy per unit
length is self-averaging and approaches a constant,
%\begin{equation}
$E(L)/L \rightarrow \langle E(L)
\rangle / L \equiv -E_0.$
%\label{eEnergyperunitlength}
%\end{equation}
Fluctuations in energy and transverse extension also
grow with length (albeit more slowly), and are described by
non-trivial power-laws
\begin{equation}
\delta E \sim A L^\omega \quad {\rm and} \quad \delta x \sim B L^\zeta,
\label{eEnergyfluctuations}
\end{equation}
where for short-range correlated disorder, the exponents
$\omega$ and $\zeta$ depend only on $d$ and obey $\omega =
2\zeta -1$\cite{Galileaninvariance}.  $\zeta$ is exactly $2/3$ in
$d=2$, approximately $0.61$ in $d=3$, and gradually reduces to 1/2 in
higher dimensions\cite{Numerics}.

First, we obtain a lower critical dimension by considering if
the delocalized FL is stable to infinitesimal pinning, $\Delta$.
If the FL is confined by the defect, following Nattermann and
Lipowsky\cite{NL} the localization length, $\ell$,
is estimated by minimizing the FL energy
\begin{equation}
E(L,\ell) = -\Delta L/\ell^{d-n} + A L/ \ell^{{1-\omega} \over \zeta}
-E_0L.
\label{eConfinement}
\end{equation}
The first term is the attractive contribution from the defect,
while the second term uses Eq.\ref{eEnergyfluctuations} to account
for the energy cost of confining the FL into $L/\ell^{1/\zeta}$
regions of length $\ell^{1/\zeta}$ and width $\ell$.
Whether or not the pinning term is dominant at large distances
determines its relevance.  A weak
potential is irrelevant for $d>d_l$, where $d_l$ is the {\em lower
critical dimension}, defined by
\begin{equation}
(d_l -n)\zeta = 1 - \omega.
\label{eIrrelevant}
\end{equation}
Using the above estimates of $\zeta$, $d_l = 2$
for columnar defects ($n=1$), while $3<d_l<4$ for
planar defects ($n=2$).  A single flux line in three
dimensions is thus always pinned by a planar defect.
When weak pinning is relevant, minimizing Eq.\ref{eConfinement} yields a
localization length that diverges for small $\Delta$ as
\begin{equation}
\ell \sim \Delta^{-\nu_{\perp}^0},
\quad {\rm where} \quad
\nu_{\perp}^0 = {\zeta \over {1-\omega - (d-n)\zeta}}.
\label{eWeakpinning}
\end{equation}
For the planar defect, $\nu_\perp^0 \approx 3.6$.
These results are also obtained by power counting in
Eq.\ref{eFreeenergy}, taking into account the rescaling of
temperature.

For $d>d_l$,
there is a transition between a delocalized phase for small
$\Delta$ and a localized phase for large $\Delta$.  At such a
transition, the localization length diverges as
\begin{equation}
\ell \sim \theta^{-\nu_\perp},
\end{equation}
where $\theta \equiv (\Delta - \Delta_c)/\Delta_c$ is the reduced
pinning strength.  This divergence is accompanied by singular behavior
in the energy,
\begin{equation}
(\langle E_s(L,\theta) \rangle+E_0 L)/L \sim \theta^{1-\alpha},
\end{equation}
which defines a ``heat capacity'' exponent $\alpha$.
For the borderline dimension $d_l=2$ for $n=1$, numerical simulations
indicate a depinning transition as the defect energy is
reduced\cite{Kreplica}.

To construct a mean field (MF) theory of delocalization,  consider
a FL very near the transition point on
the localized side.  Such a line makes large excursions away from the
defect, forming ``bubbles'' of typical size $z$ and $\ell\sim z^\zeta$
in the longitudinal and transverse directions (see Fig.\ \ref{fig1}).
For large $z$, the energy of a single bubble relative to a pinned
segment is
\begin{equation}
{\cal E}(z) = (\Delta-E_0)z - Az^\omega.
\label{eBubbleenergy}
\end{equation}
The first term is the energy cost (per unit length) of leaving the
defect to wander in the bulk; the second term is a typical
excess energy gain available (at scale $z$) due to a favorable
arrangement of impurities.  Minimizing this
energy gives the scaling $z \sim
\theta^{-\nu_{\parallel}}$ and $\ell \sim \theta^{-\nu_{\perp}}$, with
\begin{equation}
\nu_\parallel(\! MF\!) = \; 1/(1-\omega), \; \;{\rm and}\; \;
\nu_\perp(\! MF\!) = \;  \zeta/(1-\omega).
\label{eScalingresultforlengthandtime}
\end{equation}
Assuming that the probability of encountering a favorable bubble
is independent of its length $z$, the total number of favorable
bubbles is proportional to $L/z$. The energy of the FL
thus scales
as
\begin{equation}
E(L) \propto -E_0 L + {L \over z}{\cal E}(z) \sim -(E_0 + \theta)L.
\label{eTotalenergy}
\end{equation}
The corresponding heat capacity exponent, $\alpha = 0$, satisfies a modified
hyperscaling form,
$1-\alpha=(1-\omega)\nu_\parallel$,
appropriate to a zero temperature fixed point.

We know of no obvious way of determining the upper critical dimension
for the validity of the MF argument.  In low dimensions, the
argument may break down in a number of ways. The energy of a bubble
configuration, in which interior returns are not allowed, may be
different from Eq.\ref{eBubbleenergy}.  Intersections of the defect
and the FL become probable when the sum of their fractal dimensions,
$n +1/\zeta$, is greater than $d$, yielding a critical dimension
$d_u$, which satisfies
\begin{equation}
(d_u - n)\zeta = 1.
\label{eUppercriticaldimension}
\end{equation}
For columnar defects, $2<d_u<3$, while for planar defects, $3<d_u<4$.
The above result is certainly a lower bound for the true upper
critical dimension, but it is important to realize that for the
thermal depinning transition it incorrectly gives $d_u-n=2$, rather
than the exact result of $d_u-n=4$.  Other factors, such as a
renormalization of the linear term in Eq.\ref{eBubbleenergy} by
smaller scale bubbles or correlations between adjacent bubbles, may
lead to a higher $d_u$ for the disorder-induced delocalization as
well.  Taken at face value, the above result implies that the case of
a columnar pin in three dimensions falls in the mean-field regime.
Application of Eq.\ref{eScalingresultforlengthandtime} then gives
$\nu_\perp \approx 0.78$, using the numerical values for $\zeta$ and
$\omega$ in $d=3$\cite{Numerics}.

We now describe two approaches that attempt to go beyond the MF
treatment. In the ``necklace model'' applied by Lipowsky and Fisher (LF)
to the borderline case of $n=1$ and $d=2$\cite{Necklace},
the FL partition function is
decomposed into configurations with all possible sequences of pinned
and unpinned (bubble) segments, and calculated by Laplace transforms.
For thermal delocalization, since the partial partition functions of
pinned and unpinned segments are known, this can be done exactly.
In the presence of randomness, the partition functions are
not known exactly, and furthermore are themselves random quantities.
LF propose using pre-averaged forms that depend only on the exponents
$\zeta$ and $\omega$.  Although it is unclear that such
pre-averaging faithfully reproduces the desired quenched quantities, LF's
results agreed with both replica\cite{Kreplica} and numerical\cite{HH}
treatments of that case.  A straightforward extension of their
treatment to general $n$ and $d$,\cite{BKunpub} gives
\begin{equation}
\label{eNecklaceresults}
\nu_\perp (NL)=
\cases{
{\zeta \over {1-(d-n)\zeta}} & for $0<(d-n)\zeta<1$ \cr
{\zeta \over {(d-n)\zeta} - 1} & for $1<(d-n)\zeta<2$ \cr
\zeta & for $2<(d-n)\zeta$.
}
\end{equation}
These results are substantially different from ours in both low and
high dimensions.
In particular, for the columnar pin in $d=3$, the necklace
model gives $\nu_\perp(NL) \approx 2.8$.
(Hwa has obtained the results of
Eq.\ref{eNecklaceresults} from a different approach\cite{Hwa}.)

A recent preprint\cite{KS} by Kolomeisky and Straley (KS) treats
this problem for $n=1$ by the renormalization group (RG), and concludes
\FL\begin{equation}
\label{eKS}
\nu_\perp (\! K\!S\!)\!\!=\!\!
\cases{
{\zeta \over {1-\omega-(d-1)\zeta}}\! & for
$\!0\!<\!(\!d\!-\!1\!)\zeta\!\!<\!1\!-\!\omega\!$ \cr
{\zeta \over {(d-1)\zeta} - 1+\omega}\! & for
$\!1\!-\!\omega\!<\!(\!d\!-\!1\!)\zeta\!\!<\!2\!-\!2\omega\!$ \cr
{\zeta\over 1-\omega}\! & for $\!2\!-\!2\omega\!<\!(\!d\!-\!1\!)\zeta\!$.
}
\end{equation}
These results coincide with ours in the first regime, and at high
dimensions. However, we remain unconvinced about the identification
of the upper critical dimension, and the exponents in the intermediate
regime\cite{comment}. For the case of the columnar pin in three
dimensions, Eq.\ref{eKS} gives $\nu_\perp(KS) \approx 1.5$.

To differentiate between these theories, we examined the problem
numerically for columnar and planar defects in three dimensions by a
transfer-matrix method, which locates the optimal path exactly in a
strip of finite width.  To enhance performance, we chose the $z$
direction along the diagonal of a cubic lattice, with random energies
on the bonds.  The energies of optimal paths terminating at position
${\bf x}$ at height $z$ obey the recursion relation
\begin{equation}
E({\bf x},z\!+\!1) \!= \!\!\! \min_{|{\bbox{{\bf x}-{\bf x'}}}| = 1}
\!\!\! \left\{ E({\bf x}'\!,\!z) +
\epsilon({\bf x},\!{\bf x}'\!,\!z) \!- \!\Delta\delta({\bf x}_\perp\!)
\right\}\!,
%\label{eMinrecurse}
\end{equation}
where $\epsilon({\bf x},{\bf x}',z)$ is a random energy for the bond
connecting ${\bf x}$ and ${\bf x}'$ at height $z$, and $\delta({\bf
x}_\perp)$ is an appropriate lattice delta function indicating when
the FL is on the defect.
 The above recursive computation
is polynomial in the length $L$,
allowing us to examine very long lines ($L = 2.5 \times 10^6$
lattice constants with a transverse width $W = 250$ lattice units).
The $z$-averaged end-point displacement from the defect settles to a
constant value,
%\begin{equation}
$\langle {\bf x}_\perp^2 \rangle_z \equiv \ell(W)^2.$
%\end{equation}
In the localized phase, $\ell(W)$ converges to a finite value
as $W \rightarrow \infty$, while in the delocalized phase, it
grows as $\ell(W)\propto W$. By examining $\ell(W)$ for several
pinning energies $\Delta$ and widths $W$, we find clear evidence for a
depinning transition from a columnar pin, but none from a planar
defect (see Fig.\ \ref{fig2}), consistent with our prediction for
$d_l$.
  Log-log fits to power-law forms yield $\nu_\perp^0 = 2.3 \pm
0.1$ for $n=2$ and $\nu_\perp = 1.3 \pm 0.6$ for $n=1$ (see Fig.\
\ref{fig3}).  Distance from the critical region may cause additional
{\sl systematic} errors.  The
curvature in Fig.\ \ref{fig3} suggests that this is indeed the case for
the planar defect, which may explain the discrepancy with the result
of Eq.\ref{eWeakpinning}.  Given the result for $n=1$, it
is tempting to rule out Eq.\ref{eNecklaceresults}.
However, we must caution that related numerical
simulations\cite{HH} in $d=2$, see an effective exponent of
$\nu_\perp=1$ (the mean-field estimate), before reaching the true
asymptotic value of $\nu_\perp=2$. Resolution of this issue requires
more extensive simulations.

The actual situation in superconductors is complicated by several
factors.  Both defects and FLs appear at finite densities and with
lengths limited by the sample thickness. First consider a single FL in
a random set of columnar pins.  The FL can be unpinned from a
particular pin by two mechanisms. When the localization length,
$\ell$, becomes comparable to the typical defect separation
$\bar{\ell}_d \equiv \rho_d^{-1/2}$, the FL can freely wander between
pins. (It may still be collectively localized by randomness in the
distribution of columnar pins at a much larger length scale.)  The
second mechanism, originally described in the context of the bose
glass\cite{HN}, involves a long-range hopping mechanism in which a FL
wanders to a distant pin. The energy cost of such a move ($\propto
L^{1/3}$) is compensated by the fluctuations in pinning energy
($\propto L^{1/2}/\ell^b$ with $b=3/(2\zeta)-2$). A long enough
sample, $L\gg L_{\text{hop}}\propto \ell^{2.8}$, will have
sufficiently many favorable regions to allow such hops.  The pinning
effects described in this paper, are thus applicable only for $L \ll
L_{\text{hop}}$ and $\ell\ll \bar{\ell}_d$.  A finite density of FLs
introduces another length scale $\bar{\ell}_{\text{FL}} \equiv
\rho_{\text{FL}}^{1/2}$. The dilute bose glass for
$\bar{\ell}_{\text{FL}}\gg \bar{\ell}_d$ is subject to the same
constraints, and is unstable to the above hopping mechanism\cite{HN}
for $L\gg L_{\text{hop}}$. In the over dense limit of
$\bar{\ell}_{\text{FL}}\ll \bar{\ell}_d$, the hopping mechanism is no
longer relevant. A remnant of the unpinning transition may still be
observed for $\ell\ll \bar{\ell}_{\text{FL}}$.  Similar considerations
hold for the case of grain boundary pinning.

Columnar pins and grain boundaries are very effective sources of FL
pinning in superconductors. However, point defects in the bulk provide
a competing mechanism for unpinning FLs from these extended defects.
We provide analytical and numerical arguments that the FL is always
pinned to a planar defect, but that there is a transition in the case
of the columnar pin.
%The numerical result for the FL liberation
%exponent, describing divergence of the localization length, is
%consistent with a ``mean-field" value of $\nu_\perp \approx 0.78$.
The requirements for observation of the
transition and the localized phase can be satisfied close to $H_{c1}$,
although the size of this region in the cuprite materials is extremely
narrow. It is hoped that more sensitive probes will be developed to
investigate these and other subtle effects in this regime.

We acknowledge discussions with T. Hwa and D. R. Nelson. This research
was supported by the NSF through grant number DMR--90--01519.

\begin{figure}
\caption{A flux line localized around a columnar pin in three
dimensions ($d=3$, $n=1$).  White portions of the FL indicate when it
is on the defect.}
\label{fig1}\end{figure}

\begin{figure}
\caption{$\ell(W)$ for various widths and defect
energies for (a) the columnar pin and (b) the planar defect.  The bond
energies were uniformly distributed integers between $0$ and $4095$.
The region of rapid change near $\Delta_c \approx 375$ indicates the
transition for the columnar pin, while no such feature is present in
the case of the planar defect.}
\label{fig2}\end{figure}

\begin{figure}
\caption{Logarithmic plots of the localization length $\ell$ versus the
reduced pinning strength for (a) the columnar pin and (b) the planar
defect.  The different curves in (a) account for the uncertain value
of $\Delta_c$. }
\label{fig3}\end{figure}

\end{document}